\documentclass[prb,showpacs,floatfix,preprint,superscriptaddress,byrevtex]{revtex4}
\usepackage{graphicx}
\usepackage{dcolumn}
\usepackage{bm}
\usepackage{amssymb}
\usepackage{amsmath}

\newcommand{\V}{V\binom{0~\mathbf{q}~-\mathbf{q}}{j~j_1~j_2}}
\newcommand{\DP}{\Phi_{\alpha\beta\gamma}\binom{0~l_2~l_3}{s_1~s_2~s_3}}

\newcommand{\mf}[1]{\mathbf{#1}}

\newcommand{\lla}{\langle\!\langle}
\newcommand{\rra}{\rangle\!\rangle}

\begin{document}
%\preprint{Draft - not for distribution}
\bibliographystyle{apsrev}
\title{Infrared Optical Properties of Ferropericlase \boldmath (Mg$_{1-x}$Fe$_x$O):
  \unboldmath Experiment and Theory}
\author{Tao Sun}
\email{tsun@grad.physics.sunysb.edu}
\author{Philip B. Allen}
\affiliation{Department of Physics and Astronomy, State University of New York,
Stony Brook, New York 11794}
\author{David G. Stahnke}
\affiliation{Department of Physics, University of California, San Diego,
La Jolla, CA 92093}
\author{Steven D. Jacobsen}
\affiliation{Department of Earth and Planetary Sciences, Northwestern
University, Evanston, IL 60208}
\author{Christopher C. Homes}
\affiliation{Condensed Matter Physics \& Materials Science Department,
Brookhaven National Laboratory, Upton, New York 11973}
\date{\today}
\begin{abstract}
The temperature dependence of the reflectance spectra of magnesium oxide (MgO)
and ferropericlase (Mg$_{1-x}$Fe$_x$O, for $x=0.06$ and $x=0.27$) have been
measured over a wide frequency range ($\approx 50$ to $32\,000$ cm$^{-1}$) at
295 and 6~K.  The complex dielectric function has been determined from a
Kramers-Kronig analysis of the reflectance.  The spectra of the doped materials
resembles pure MgO in the infrared region, but with much broader resonances. We
use a shell model to calculate the dielectric function of ferropericlase,
including both anharmonic phonon-phonon interactions and disorder scattering.
These data are relevant to understanding the heat conductivity of
ferropericlase in the earth's lower mantle.
\end{abstract}
%
% PACS
%
%  63.20.-e Phonons in crystal lattices
%  63.20.Dj Phonons, normal modes, phonon dispersion
%  63.20.Kr Phonon-electron and phonon-phonon interactions
%  63.20.+m Phonons in low-dimensional structures
%  63.20.Mt Phonon-defect interactions
%  71.15.Mb Density functional theory, LDA, gradient and other corrections
%  71.45.Lr CDW systems
%  74.25.Gz Optical properties
%  74.25.Kc Phonons
%  78.30.-j Infrared and Raman spectra
%
\maketitle
%

%
% Introduction
%
\section{Introduction}
Ferropericlase, (Mg$_{1-x}$Fe$_x$O, with $x=0.10$-$0.15$), is thought to be
one of the major constituents of the earth's lower
mantle~($660$-$2900$ km depth).\cite{mantle} The name `magnesiow\"{u}stite' 
is also used, but properly refers to the doping region x close to the 
w\"{u}stite~(x=1), rather than the periclase~(x=0) limit. Transport properties 
of (Mg,Fe)O are therefore important in modeling the Earth's thermal state and 
evolution, where both conduction and convection are 
operative.\cite{hofmeister99, hofmeister05} The two heat carriers in 
conduction for an insulating mineral like ferropericlase are phonons and 
photons. Phonons~(extended or localized) are distributed in the far and 
mid-infrared frequency range. They can be 
scattered by various defects~(e.g. impurities, grain boundaries, \dots) and by 
the intrinsic anharmonic phonon-phonon interactions. Photons are described by
Planck's black-body radiation formula, reaching energies $\sim 1$~eV at
$2000$-$3000$~K. The photon-matter interaction is more complex. Far- and
mid-infrared photons usually couple with infrared-active vibrations. Photons
of higher frequency can induce pure electronic transitions or vibronic
transitions, depending on the details of the system's electronic states and
adiabatic potential-energy surfaces. Goncharov {\it et al.}\cite{goncharov}
measured the optical absorption spectra of Mg$_{1-x}$Fe$_x$O~(with x=0.06,
0.15, and 0.25) across the high-spin/low-spin
transition,\cite{sherman, badro, lin} which occurs over
a pressure range of 40-60 GPa at room temperature. Their results indicate that
low-spin~(Mg,Fe)O will exhibit lower~(rather than higher\cite{burns})
radiative heat conductivity than high-spin phase due to the red-shift of the
charge-transfer edge. The origin of this spin transition and its influence on
the radiative heat conductivity of ferropericlase are further investigated
recently.\cite{keppler} A complete picture of the thermal conductivity must
include contributions from both phonons and photons.

As a solid solution, ferropericlase has a vibrational frequency spectrum
similar to that of pure MgO. However, with strong disorder scattering of
propagating vibrational states, the harmonic eigenstates of the disordered
crystal do not necessarily have a well defined wave number, and may not
propagate ballistically. In addition, the anharmonic phonon-phonon
interactions causes a shoulder at $\sim 640$~cm$^{-1}$ in the infrared~(IR)
reflectance spectrum of pure MgO.\cite{jasperse} Thus anharmonicity should
also be included in analyzing the infrared reflectance of ferropericlase.

In this paper we report the temperature-dependent infrared reflectance
measurements of magnesium oxide and ferropericlase  for several Fe
concentrations at ambient pressure. We construct a model in which anharmonic
phonon-phonon interactions and disorder scattering are treated separately.
Their effects are then combined for comparison with the experimental data.

%
% Experimental
%
\section{Experimental Measurements}
\begin{figure*}
\includegraphics[width=0.96\textwidth]{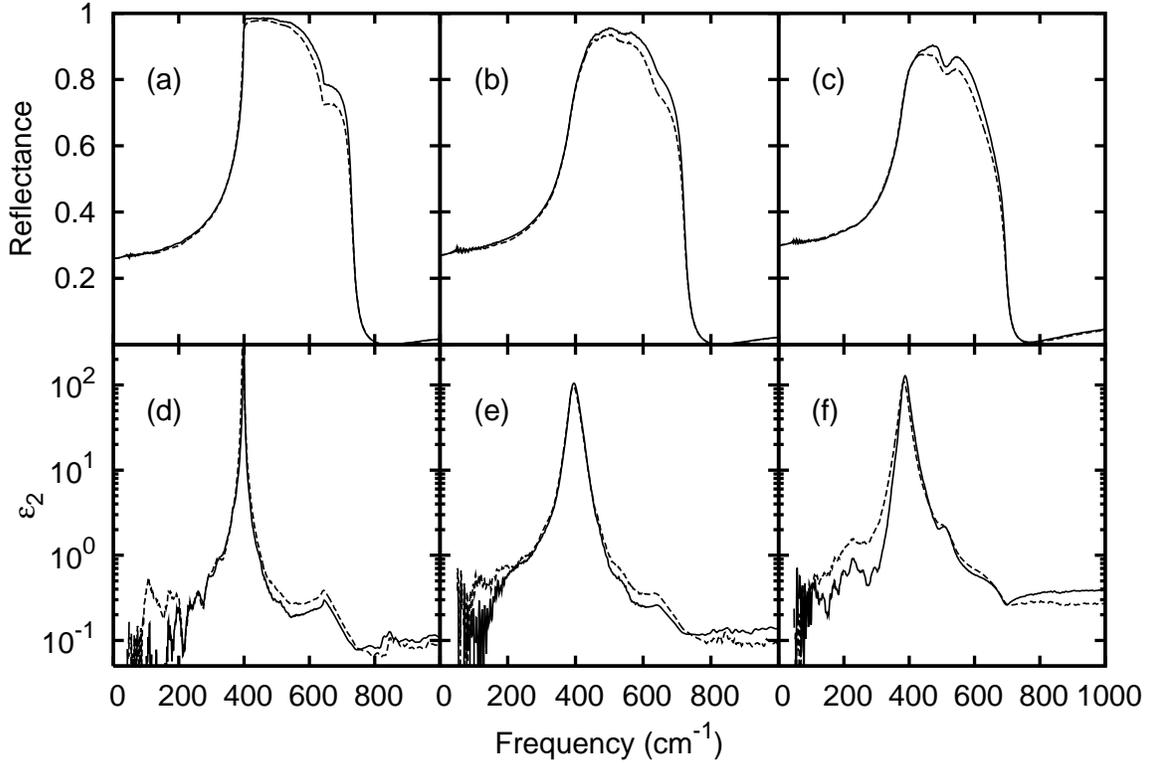}
\caption{The measured reflectance $R(\omega)$; (a) pure MgO, and
Mg$_{1-x}$Fe$_x$O for (b) 6\% and (c) 27\% Fe-doping. The corresponding
imaginary part of the dielectric functions $\epsilon_2 (\omega)$; (d) pure MgO,
and Mg$_{1-x}$Fe$_x$O for (e) 6\% and (f) 27\% Fe-doping. The solid line
corresponds to data measured at 6~K, dashed line corresponds to data at 295~K. }%
\label{fig:exp}
\end{figure*}

The samples we examined are homogeneously doped single crystals, in which
Fe$^{3+}/\sum {\rm Fe}\approx 0.02$ for the 6\% sample and 0.05 for the 27\%
sample. However, in our analysis the influence of Fe$^{3+}$ and magnesium
vacancies is ignored. A detailed description on the samples' synthesis,
crystallography and elastic properties is in Ref.~\onlinecite{jacobsen}. The
samples are rectangular slabs with typical dimensions of $1\,{\rm mm} \times
2\,{\rm mm}$, with a thickness of $\approx 0.3$~mm.  In order to reduce
interference effects due to reflections from the back surface, the samples have
been wedged.  However, due to the thin nature of the samples, the largest wedge
that could be introduced was $\simeq 15^\circ$.
The reflectance spectra has been measured at a near-normal angle of incidence
at 295 and 6~K over a wide frequency range from $\approx 50$ to about
$32\,000$~cm$^{-1}$ on Bruker IFS 66v/S and 113v spectrometers using an {\it
in-situ} evaporation technique.\cite{homes93}  The measured reflectance at 295
and 6~K of pure MgO, and Mg$_{1-x}$Fe$_x$O, for $x=0.06$ and $x=0.27$ are shown
in Figs.~\ref{fig:exp}(a), (b) and (c), respectively.  Although wedging the
samples has been very effective at reducing interference effects, weak fringes
may still be detected at low temperature below about 150~cm$^{-1}$.
The complex dielectric function $\epsilon = \epsilon_1 +i\epsilon_2$ has been
determined from a Kramers-Kronig analysis of the reflectance, where
extrapolations are supplied for $\omega \rightarrow 0, \infty$.  At low
frequency, an insulating response is assumed and $R(\omega \rightarrow 0)
\simeq 0.27$, 0.28 and 0.31 for MgO, and the 6\% and 27\% Fe-doped materials,
respectively.  Above the highest measured frequency the reflectance has been
assumed to be constant to approximately $75\,000$~cm$^{-1}$, above which a
free-electron approximation has been assumed $(R\propto 1/\omega^4)$. The
imaginary part of the resulting dielectric function at 6 and 295~K of pure MgO,
and Mg$_{1-x}$Fe$_x$O, for $x=0.06$ and $x=0.27$, are shown in
Figs.~\ref{fig:exp}(d), (e) and (f), respectively.  The imaginary part of the
dielectric function contains most of the physical information, and is the focus
of our theoretical analysis. The optical features have been fit to a classical
oscillator model using the complex dielectric function
\begin{equation}
%  \tilde\epsilon(\omega) = \epsilon_\infty + \sum_j {{\omega_{p,j}^2} \over
   \epsilon(\omega) = \epsilon_\infty + \sum_j \frac{\omega_{p,j}^2}
  {\omega_{{\rm TO},j}^2-\omega^2-i2\omega\gamma_j},
  \label{equa:classical}
\end{equation}
where $\epsilon_\infty$ is a high-frequency contribution, and $\omega_{{\rm
TO},j}$, $2\gamma_j$ and $\omega_{p,j}$ are the frequency, full width and 
effective plasma frequency of the $j$th vibration. The results of non-linear
least-squares fits to the reflectance and $\epsilon_2(\omega)$ are shown in
Table~\ref{tab:exp}. In addition to the strong feature in $\epsilon_2(\omega)$
seen at about 400~cm$^{-1}$, other features at $\approx 520$ and $\approx
640$~cm$^{-1}$ are also clearly visible in $\epsilon_2(\omega)$ shown in
Fig.~\ref{fig:exp}; however, these features are very weak and as a result the
the strengths and widths of these modes are difficult to determine reliably.

%
% Figure 1
%

%
% Table I
%
\begin{table*}
\caption{A comparison of the fitted values of the static and high-frequency
contributions to the real part of the dielectric function at room temperature,
as well as the fitted frequency, full width and effective plasma frequency
($\omega_{\rm TO}$, $2\gamma$ and $\omega_{p}$, respectively) of the feature
associated with the strong TO mode in MgO, and the 6\% and 27\% Fe-doped
materials at 295 and 6~K. The units of $\omega_{\rm TO}$, $2\gamma$ and
$\omega_{p}$ are in cm$^{-1}$.
The strength of the TO mode is also expressed as a dimensionless oscillator
strength $S=\omega_{p}^{2}/\omega_{\rm TO}^{2}$. }
\begin{ruledtabular}
\begin{tabular}{ccccccccc}
 & & & & 295~K & & & 6~K & \\
 Mg$_{1-x}$Fe$_x$O & $\epsilon_0^a$ & $\epsilon_\infty^a$ &
   $\omega_{\rm TO}^b$ & $2\gamma^c$ & $\omega_{p}^d \ \ (S)$ &
   $\omega_{\rm TO}^b$ & $2\gamma^c$ & $\omega_{p}^d \ \ (S)$\\
\cline{1-1} \cline{2-3} \cline{4-6} \cline{7-9}
 pure     &  9.2 & 2.95 & 396.5 & 3.44 & 1010~(6.5) & 398.9 & 1.72 & 1030~(6.7) \\
 $x=0.06$ & 10.8 & 3.10 & 395.6 & 30.5  & 1090~(7.6) & 396.7 & 29.1 & 1120~(8.0) \\
 $x=0.27$ & 11.8 & 3.65 & 384.5 & 28.6  & 1100~(8.2) & 388.6 & 25.7 & 1140~(8.6) \\
\end{tabular}
\end{ruledtabular}
\footnotetext[1]{Values at 295~K, the estimated uncertainty is about $\pm
0.1$.}
\footnotetext[2]{The uncertainty in $\omega_{\rm TO}$ is $\pm 0.1$~cm$^{-1}$.}
\footnotetext[3]{The uncertainties for $2\gamma$ are $\pm 0.1$~cm$^{-1}$ in the
pure material, and $\pm 0.5$~cm$^{-1}$ in the Fe-doped materials. }
\footnotetext[4]{The uncertainty in $\omega_p$ is $\pm 20$~cm$^{-1}$. }
\label{tab:exp}
\end{table*}

%
% Computational methods
%
\section{Computational Methods}
\subsection{General Scheme}
Infrared dielectric properties of ionic crystals are contained in the linear
response function $\epsilon_{\alpha\beta}(\omega)= \epsilon_{\alpha\beta} (
\infty) + 4\pi\chi_{\alpha\beta}(\omega)$.\cite{cowley} Considering only the
first-order moment of the electric dipole, the dielectric susceptibility of a
crystal can be related to its displacement-displacement retarded Green's
function by:
\begin{eqnarray}
\chi_{\alpha\beta}(\omega)&=&-\frac{1}{NV_c}\int_{-\infty}^{+\infty}
\frac{\theta(t-t')}{i\hbar}\langle[D_{\alpha}(t),D_{\beta}(t')]\rangle
e^{i\omega(t-t')}\,d(t-t')\nonumber \\
&=&-\frac{1}{NV_c}\sum_{ls\gamma}\sum_{l's'\delta}Z_{\alpha \gamma}(ls)Z_{\beta
\delta}(l's')\int_{-\infty}^{+\infty}\frac{\theta(t-t')}
{i\hbar}\langle[u_{\gamma}(ls;t),u_{\delta}(l's';t')]\rangle
e^{i\omega(t-t')}\,d(t-t') \nonumber \\
&=&-\frac{1}{NV_c}\sum_{ls\gamma}\sum_{l's'\delta}
Z_{\alpha\gamma}(ls)Z_{\beta\delta}(l's')G_{\gamma\delta}(ls,l's';\omega),
\label{equa:susceptibility}
\end{eqnarray}
where
$D_{\alpha}(t)=\displaystyle\sum_{ls\beta}Z_{\alpha\beta}(ls)u_{\beta}(ls;t)$
is the $\alpha$ component of the first order electric dipole of the whole
crystal, $Z_{\alpha\beta}(ls)$ is the Born effective charge tensor of the atom
$s$ at site $l$, and $u_{\beta}(ls;t)$ is the atom's displacement at time $t$.
The volume of a single cell is $V_c$, and $N$ is the number of the cells in the
whole crystal. The Green's function $G_{\alpha\beta}(ls,l's';t-t')$ is defined
as:
\begin{eqnarray}
G_{\alpha\beta}(ls,l's';t-t')=\frac{\theta(t-t')}{i\hbar}
\langle[u_{\alpha}(ls;t),u_{\beta}(l's';t')]
\rangle,
\label{equa:retgreen}
\end{eqnarray}
which can be evaluated from its equation of motion.\cite{taylor} For a harmonic
crystal, the vibrational Hamiltonian is quadratic and can be solved exactly. We
denote the eigenvectors of a pure crystal as
$\frac{1}{\sqrt{N}}\hat{e}_{\alpha}(s|\mf{q}j)e^{i\mf{q}\cdot\mf{R}(ls)}$, the
corresponding eigenvalues as $\omega_{\mf{q}j}$, those of a disordered crystal 
as $e_{\alpha}(s|j)$ and $\omega_{j}$, the Green's function of the pure as 
$\mf{g}$, the disordered as $\mf{G}^{0}$. Then
\begin{eqnarray}
g_{\alpha\beta}(ls,l's';\omega)&=&
\sum_{\mf{q}j}\frac{\hat{e}_{\alpha}(s|\mf{q}j)\hat{e}^{*}_{\beta}(s'|\mf{q}j)
e^{i\mf{q}\cdot(\mf{R}(ls)-\mf{R}(l's'))}}
{N\sqrt{M(s)M(s')}(\omega^{2}-\omega_{\mf{q}j}^2+i2\omega\eta)},
\label{equa:geigen} \\
G^{0}_{\alpha\beta}(ls,l's';\omega)&=&
\sum_{j}\frac{e_{\alpha}(ls|j)e^{*}_{\beta}(l's'|j)}
{\sqrt{M(ls)M(l's')}(\omega^{2}-\omega_{j}^2+i2\omega\eta)},
\label{equa:Geigen}
\end{eqnarray}
where the mass of the atom $s$ is denoted as $M(s)$ in the pure crystal, 
$M(ls)$ in the disordered crystal, with the extra label $l$ to specify its site,
$\eta$ is an infinitesimal number ensuring causality.

Anharmonic interaction will couple these modes and make exact solution
impossible. The standard treatment of this many-body effect uses the Dyson
equation to define a self-energy for each mode. We can either choose
$e_{\alpha}(s|j)$ as the unperturbed states, then the only interaction will be
anharmonicity, or choose $\frac{1}{\sqrt{N}}\hat{e}_{\alpha}
(s|\mf{q}j)e^{i\mf{q}\cdot\mf{R}(ls)}$ as the basis and treat disorder as an
extra perturbation. The first approach has been used by one of the
authors~(PBA) to study the anharmonic decay of vibrational states in amorphous
silicon. \cite{fabian} In this paper we use a hybrid approach. We write the
dielectric function of a disordered anharmonic crystal in the perfect crystal
harmonic basis as 
\begin{eqnarray}
\epsilon_{\alpha\beta}(\omega)&=&
\epsilon_{\alpha\beta}(\infty)+4\pi\chi_{\alpha\beta}(\omega)\nonumber\\
&=&\epsilon_{\alpha\beta}(\infty)+ \frac{4\pi}{V_c}\sum_{j=1}^{\rm TO}
\frac{\displaystyle\sum_{s\gamma}Z_{\alpha\gamma}(s)
\frac{\hat{e}_{\gamma}(s|0j)}{\sqrt{M(s)}}
\displaystyle\sum_{s'\delta}Z_{\beta\delta}(s')
\frac{\hat{e}^{*}_{\delta}(s'|0j)}{\sqrt{M(s')}}}
{\{\omega(0j)^2-\omega^2+2\omega(0j)(\Delta(0j,\omega)-i\Gamma(0j,\omega)\}},
\label{equa:eps-green}
\end{eqnarray}
where $\omega(0j)\equiv\omega_{{\rm TO},j}$ is the frequency at $\mf{q}=0$ of
the $j$th TO branch.  The terms $\Delta(0j,\omega)$ and $\Gamma(0j,\omega)$
correspond to the real and imaginary part of the mode's self-energy
$\Sigma(0j,\omega)$. Then we split this self-energy into two parts: 
$\Sigma = \Sigma_{\rm anharm}+\Sigma_{\rm disorder}$. 
Each piece is calculated independently. This is equivalent to omitting all the 
diagrams where the disorder scattering vertex appears inside an anharmonic 
interaction loop. The anharmonic interaction of ferropericlase is assumed to be 
the same as that of pure MgO, i.e. the influence of disorder on anharmonic 
coefficients is totally ignored. Disorder is treated by exact diagonalization 
without anharmonicity, it is then converted to a self-energy of the TO mode in 
the perfect crystal harmonic basis. These approximations are tested by 
comparing with the experimental results.

%
% The shell model
%
\subsection{Shell Model}
The scheme described above is general. It does not depend on which microscopic
model is chosen to get harmonic phonons, disorder scattering strength, and
higher-order force constants. Here we use an anharmonic shell model, with shell
parameters fitted to experiments. The harmonic phonon properties in this paper
are calculated with the general utility lattice program (GULP) code.\cite{gulp}

\begin{figure}
\includegraphics[width=0.48\textwidth]{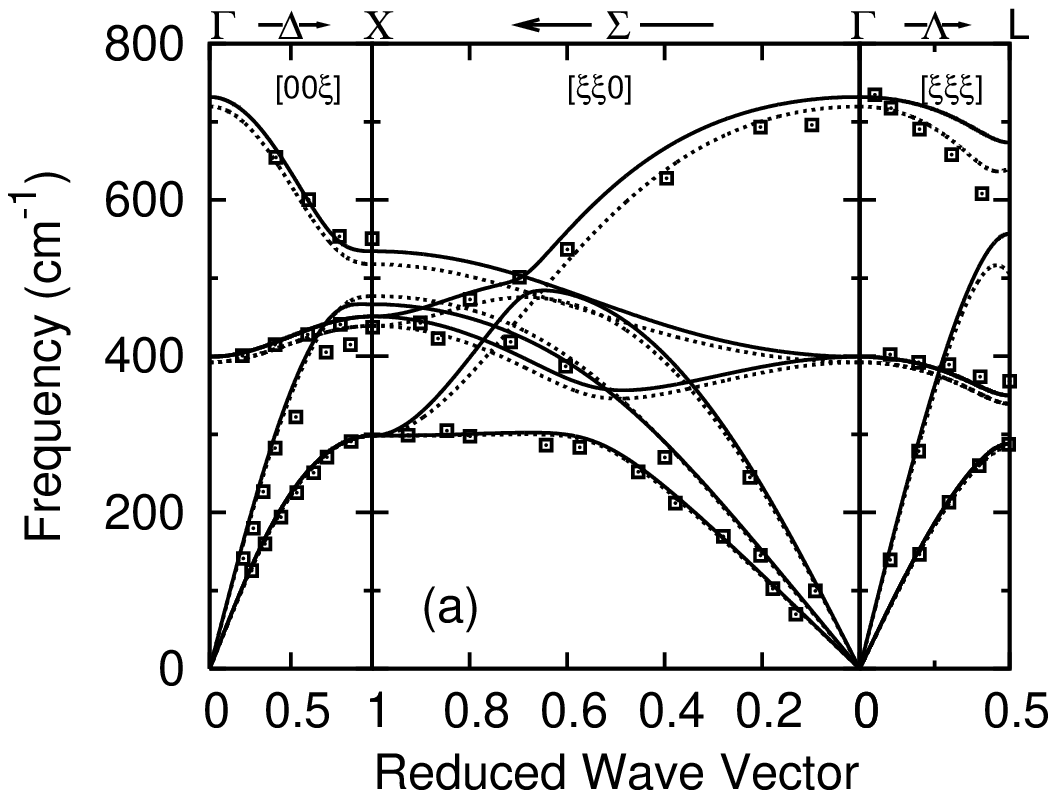}
\includegraphics[width=0.48\textwidth]{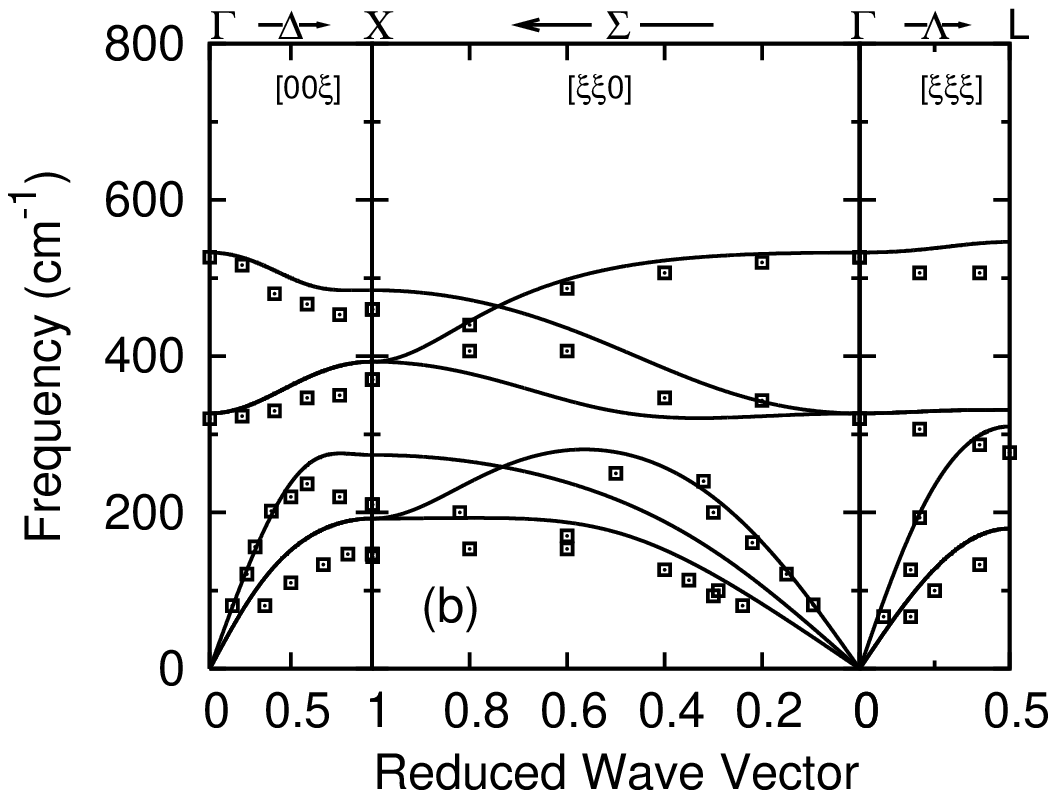}
\caption{Phonon dispersions of the pure crystals.
(a)~MgO, solid line corresponds to the rigid-shell model S-I, dashed line to
the isotropic breathing-shell model B, dots are the experimental data taken
from Ref.~\onlinecite{sangster};
(b)~FeO, solid line corresponds to the rigid-shell model S-II, dots are the
experimental data taken from Ref.~\onlinecite{kugel}.}
\label{fig:phonon}
\end{figure}

Two sets of shell parameters\cite{gulp,poten} are used for MgO: S-I and B, and
one for FeO: S-II. S-I and S-II are rigid shell models in which O$^{2-}$ has
the same set of parameters, thus they can be conveniently used to simulate
ferropericlase. B is an isotropic breathing shell model which gives better fit
to the experimental data. However, it can not be directly used for
ferropericlase. For FeO the elastic constants C$_{12}>$C$_{44}$, while the
isotropic breathing shell model is only suitable for cases where
C$_{12}<$C$_{44}$.\cite{sangster2} We treat B as a reference to check our
anharmonic calculations based on S-I. All the model parameters are listed in
Table~\ref{tab:shelpara}. Table~\ref{tab:pure} contains the calculated physical
properties and corresponding experimental values. Phonon dispersion curves for
the pure crystals of MgO and FeO are shown in Fig.~\ref{fig:phonon}.

%
% Figure 2
%

%
% Table II
%
\begin{table}
\caption{Shell model parameters used in the calculation.\cite{gulp,poten} The
short-range repulsive potential is assumed to be a two-body Buckingham type:
for S-I and S-II, $V(r) = A\exp(-r/\rho)-C/r^6$; for B, $V(r)
 = A\exp(-(r-r_{0})/\rho) - C/r^6$. The parameter k represents the spring constant
between core and shell. Rows in which atomic symbols have a star (*) are for
the B model. The label `shell' denotes a potential that acts on the central
position of the shell, while `bshell' denotes an interaction that acts on the
radius of the shell which was fixed at 1.2~\AA. An extra parameter in B model
is k$_{\rm BSM} = 351.439$~eV\AA$^{-2}$. The equilibrium shell radius $r_{0}$
is 1.1315~{\AA}\ after optimization.}
\begin{ruledtabular}
\begin{tabular}{cccc}
&Z$_{\rm core}$ (e)&Z$_{\rm shell}$ (e) &k (eV) \\
\hline
O&$0.9345$&$-2.9345$&$51.712$\\
Mg&$2$&$-$&$-$\\
Fe&$-1.1682$&$3.1682$&$69.562$\\
O$^{*}$&$0.8$&$-2.8$&$46.1524$\\
Mg$^{*}$&$2$&$-$&$-$ \\
\hline
& A (eV) &$\rho$ (\AA) & C (eV$\cdot${\AA}$^6$) \\
\hline
O shell-O shell&$22764.3$&$0.149$&$20.37$\\
Mg core-O shell&$1346.6$&$0.2984$&$0.0$\\
Fe shell-O shell&$1231.2$&$0.3086$&$0.0$\\
O$^{*}$ shell-O$^{*}$ shell&0.0&0.3&54.038\\
Mg$^{*}$ core-O$^{*}$ bshell&28.7374&0.3092&0.0\\
\end{tabular}
\end{ruledtabular}
\label{tab:shelpara}
\end{table}

%
% Table III
%
\begin{table}
\caption{Physical properties of pure MgO and FeO, compared with shell model
results.}
\begin{ruledtabular}
\begin{tabular}{cccccccc}
&a~(\AA)&C$_{11}$~(GPa)&C$_{12}$~(GPa)&C$_{44}$~(GPa)&$\epsilon_{0}$
&$\epsilon_{\infty}$&TO~(cm$^{-1}$)\\
\hline
MgO~(exp\cite{gulp,pold,poten})&4.212&297.0&95.2&155.7&9.86&2.96&401\\
S-I&4.225&370.9&163.0&163.0&9.88&2.94&399\\
B&4.212&297.0&95.0&155.7&9.89&2.94&392\\
\hline
FeO~(exp\cite{pold})&4.310&359&156&56&14.2&5.4&320\\
S-II&4.324&327&149&149&14.18&5.34&327
\end{tabular}
\end{ruledtabular}
\label{tab:pure}
\end{table}

%
% Anharmonicity
%
\subsection{Anharmonicity}
A complete calculation of anharmonicity is tedious, even for a pure
crystal.\cite{cowley} Thus we ignore the less important terms and focus on the
dominant one. From Eq.~(\ref{equa:eps-green}) it is clear that since
$|\Sigma|=|\Delta-i\Gamma|$ is small compared to $\omega_{\rm TO}$, the real
part of the self-energy $\Delta$ has negligible influence on
$\epsilon_2(\omega)$, except to shift its resonant frequency. The shell models
we use are fitted to the experimental data measured at room temperature. The
anharmonic shift is small, compared with the shift caused by impurity
scattering. Thus, we ignore it completely and only consider the imaginary part
of the self energy $\Gamma_{\rm anharm}(0j,\omega)$. To the lowest order
$\Gamma_{\rm anharm}(0j,\omega)$ can be written as\cite{cowley}
\begin{eqnarray}
\Gamma_{\rm anharm}(0j,\omega)&=&\frac{18\pi}{\hbar^2}\sum_{\mf{q} j_1
j_2}\left|\V\right|^2\{(n_1+n_2+1)[\delta(\omega_1+\omega_2-\omega)-\delta(\omega_1+\omega_2+\omega)]\nonumber\\
&&+(n_2-n_1)[\delta(\omega_2-\omega_1+\omega)-\delta(\omega_1-\omega_2+\omega)]\},
\label{equa:gamma}
\end{eqnarray}
where $n_1=n(\mf{q}j_1)$ is the Bose-Einsein population factor of the mode,
and $\omega_1=\omega(\mf{q}j_1)$ is the corresponding frequency. The
anharmonic coefficient $\V$ is
\begin{eqnarray}
\V=\frac{1}{3!}\sum_{\alpha\beta\gamma}\sum_{s_1 s_2 s_3}\sum_{l_2
l_3}\DP\hat{e}_{\alpha}(s_1|0j)\hat{e}_{\beta}(s_2|\mf{q}j_1)
\hat{e}_{\gamma}(s_3|-\mf{q}j_2)\nonumber\\
\left(\frac{\hbar^3}{8\omega(0j)\omega(\mf{q}j_1)\omega(-\mf{q}j_2)
M(s_1)M(s_2)M(s_3)}\right)^{\frac{1}{2}}\exp\left\{i
\mf{q}\cdot\left[\mf{R}(l_2 s_2)-\mf{R}(l_3 s_3)\right]\right\}.
\label{equa:anharm}
\end{eqnarray}
The third-order force constants $\DP$ are large only for the nearest neighbors.
Symmetry will restrict most of them to be zero, and among those nonzero terms
only two are independent. The general formula for third order force constants
is\cite{cowley}
\begin{eqnarray}
\Phi_{\alpha\beta\gamma}\binom{0~0~l'}{s~s~s'}&=&BR_{\alpha}R_{\beta}R_{\gamma}
+C(R_{\alpha}\delta_{\beta\gamma}+R_{\beta}\delta_{\alpha\gamma}+R_{\gamma}
\delta_{\alpha\beta}), \nonumber \\
B&=&\frac{\phi{'''}}{R^3}-\frac{3\phi{''}}{R^4}+\frac{3\phi{'}}{R^5},
\nonumber \\
C&=&\frac{\phi{''}}{R^2}-\frac{\phi{'}}{R^3},
\label{equa:thirdforce}
\end{eqnarray}
where $R$ is the lattice distance between the ion $\binom{0}{s}$ and
$\binom{l'}{s'}$, and $R_{\alpha}$ is its projection along $\alpha$ direction.
The term $\phi(r)$ is the two-body pair potential, and $\phi'$, $\phi'' \dots$
are derivatives with respect to $r$. Following E.~R.~Cowley,\cite{ercowley} we
compute $\Phi_{\alpha\beta\gamma}\binom{0~0~l'}{s~s~s'}$ by direct
differentiation over the nearest-neighbour short-range potentials and Coulomb
potentials. For the rigid-shell model S-I,
$\phi(r)=A\exp(-r/\rho)-\frac{4e^2}{r}$. For the breathing-shell model B,
$\phi(r)=A\exp(-(r-r_{0})/\rho)-\frac{4e^2}{r}$. To be more specific, if we
take a Mg$^{2+}$ as the origin and denote it as 1, its nearest neighbor
O$^{2-}$ along the [100] direction as 2, then from symmetry we can determine
$\Phi_{\rm xxx}(112)=\Phi_{\rm xxx}(121)=-\Phi_{\rm xxx}(122)
=\Phi_{\rm yyy}(112)=\cdots$,
$\Phi_{\rm xyy}(112)=\Phi_{\rm xzz}(112)=\cdots$.
Putting in numbers from Table~\ref{tab:shelpara} we obtain $\Phi_{\rm
xxx}(112)=-25.34$~eV/\AA$^3$, $\Phi_{\rm xyy}(112)=-1.79$~eV/\AA$^3$ for the
S-I model, and $\Phi_{\rm xxx}(112)=-24.0$~eV/\AA$^3$, $\Phi_{\rm xyy}(112)=
-1.78$~eV/\AA$^3$ for the B model. If we do not include the Coulomb
interaction, these values will be $\Phi_{\rm xxx}(112)=-42.70$~eV/\AA$^3$
$\Phi_{\rm xyy}(112)=6.88$~eV/\AA$^3$ for the S-I model, and $\Phi_{\rm
xxx}(112)= -41.57$~eV/\AA$^3$, $\Phi_{\rm xyy}(112)=7$~eV/\AA$^3$ for the B
model. Although $\Phi_{\rm xyy}(112)$ is small compared with $\Phi_{\rm
xxx}(112)$, it can still have non-negligible influence on the amplitude of
the $\Gamma(0j,\omega)$ near 640 cm$^{-1}$. Other parameters~(Born effective 
charge tensors, harmonic eigenvectors) are obtained from GULP. The integration 
over q-space is done with the tetrahedron method, using 1/48 of the Brillouin 
zone, and averaging over x, y, and z polarizations. We use 3345 q-points, 
equivalent to 160560 q-points in the whole Brillouin zone.

%
% Disorder scattering
%
\subsection{Disorder Scattering}
The self-energy of a vibrational mode caused by disorder scattering is defined
statistically,\cite{taylor,elliott} 
\begin{eqnarray}
\lla\mf{G^{0}}\rra
=\mf{g}+\mf{g}\mf{\Sigma}\lla\mf{G^{0}}\rra, \nonumber
\end{eqnarray}
where $\lla\mf{G^{0}}\rra$ denotes the Green's function averaged over different
impurity distributions. We slightly modify this definition by including the 
Born effective charge. From Eqs.~(\ref{equa:geigen}) and (\ref{equa:Geigen}), 
we define the following equation
\begin{eqnarray}
\lla\sum_{\gamma}\sum_{\delta}
Z_{\alpha\gamma}(ls)G^{0}_{\gamma\delta}(ls,l's';\omega)
Z_{\delta\beta}(l's')\rra=
\sum_{\gamma}\sum_{\delta}Z_{\alpha\gamma}(s)\tilde{g}_{\gamma\delta}
(ls,l's';\omega)Z_{\delta\beta}(s'),
\label{equa:g2G}
\end{eqnarray}
where
\begin{eqnarray}
\tilde{g}_{\gamma\delta}(ls,l's';\omega)=
\sum_{\mf{q}j}\frac{\hat{e}_{\gamma}(s|\mf{q}j)\hat{e}^{*}_{\delta}(s'|\mf{q}j)
e^{i\mf{q}\cdot(\mf{R}(ls)-\mf{R}(l's'))}}
{N\sqrt{M(s)M(s')}(\omega^{2}-\omega_{\mf{q}j}^2-i2\omega_{\mf{q}j}
\Sigma_{\rm disorder}(\mf{q}j,\omega))}.
\label{equa:gtilde}
\end{eqnarray}

The self-energy defined in this way guarantees that the dielectric
susceptibilities calculated from $\mf{G}^{0}$ and $\mf{\tilde{g}}$ are the
same. Summing over all sites of the crystal leaves only TO modes on the right
hand side of Eq.~(\ref{equa:g2G}). Thus, once we get the averaged dielectric
susceptibility $\lla\chi_{\alpha\beta}\rra$ from the exact eigenvectors of the
disordered crystal, we can extract the self-energy of its TO phonon.

We expand an orthogonal 8-atom MgO unit cell in each direction by 5 times,
which gives a $5\times5\times5$ super-cell containing 1000 atoms. Then we
randomly replace the corresponding number of Mg$^{2+}$ by Fe$^{2+}$. The shell
parameters of Mg$^{2+}$ are from S-I model, those of Fe$^{2+}$ are from S-II
model, those of O$^{2-}$ are the same in both models. From
Eqs.~(\ref{equa:susceptibility}) and (\ref{equa:Geigen}), for each
configuration we have a harmonic susceptibility
\begin{eqnarray}
\chi_{\alpha\beta}(\omega)= \frac{1}{NV_c} \sum_{j=1}^{modes}
\frac{\displaystyle\sum_{ls\gamma}Z_{\alpha\gamma}(ls)
\frac{e_{\gamma}(ls|j)}{\sqrt{M(ls)}}
\displaystyle\sum_{l's'\delta}Z_{\beta\delta}(l's')
\frac{e^{*}_{\delta}(l's'|j)}{\sqrt{M(l's')}}}
{\omega_{j}^2-\omega^2-i2\omega\eta}.
\label{equa:ave_susce}
\end{eqnarray}
We can choose a small value for $\eta$ and evaluate 
Eq.~(\ref{equa:ave_susce}) directly~(Lorentzian broadening). However, insofar
as $\eta$ is finite, it is equivalent to have each mode $j$ in 
Eq.~(\ref{equa:ave_susce}) an imaginary self-energy~(life time) 
linear in frequency $\omega$. The self-energy of the TO phonon 
$\Sigma_{\rm disorder}$ extracted from this approach will depend on frequency 
linearly. Replacing the factor $2\omega$ by $2\omega_j$ won't help either, 
as each mode $j$ now has a life time independent of frequency, and 
$\Sigma_{\rm disorder}$ will be a constant depending on $\eta$ when 
$\omega\rightarrow 0$. To avoid such artifacts we use
\begin{eqnarray}
\frac{1}{\omega_j^2-\omega^2-i2\omega\eta}=\frac{1}{\omega_j^2-\omega^2}
+\frac{i\pi}{2\omega}\left[\delta(\omega-\omega_j)+\delta(\omega+\omega_j)
\right] \nonumber
\label{equa:delta}
\end{eqnarray}
to separate the real~($\chi_1$) and imaginary part~($\chi_2$) of the dielectric 
susceptibility. Then we divide the vibrational spectrum into equally sized
bins~(1~cm$^{-1}$) and compute $\chi_2$ as a histogram. The real part $\chi_1$ 
is obtained from $\chi_2$ from the Kramers-Kronig relation. Many such 
super-cells are built and their $\epsilon_{\infty}$ and $\chi$ calculated. We 
find that 10 configurations are sufficient to give a well converged average. 
The final $\epsilon_{\infty}$ and $\chi$ are assumed to be the averaged values 
of all configurations. To remove the unphysical spikes caused by the finite 
size of our super-cells, while keeping the main features unchanged, we further 
smooth the dielectric susceptibility by averaging over adjacent bins 
iteratively,
\begin{equation}
\chi_2^{n+1}(j)=\frac{1}{6}\left[\chi_2^{n}(j-1)+4\chi_2^{n}(j)+
\chi_2^{n}(j+1)\right].
\label{equa:eps-ave}
\end{equation}
In this way we successfully simulate the dielectric function of a `real'
crystal~(real in the sense that except for finite size, disorder scattering is
treated without any further approximations). These results, together with
anharmonicity, are summarized in the next section.

%
% Comparisons and Discussion
%
\section{Comparisons and Discussion}
\begin{figure}
\includegraphics[width=0.48\textwidth]{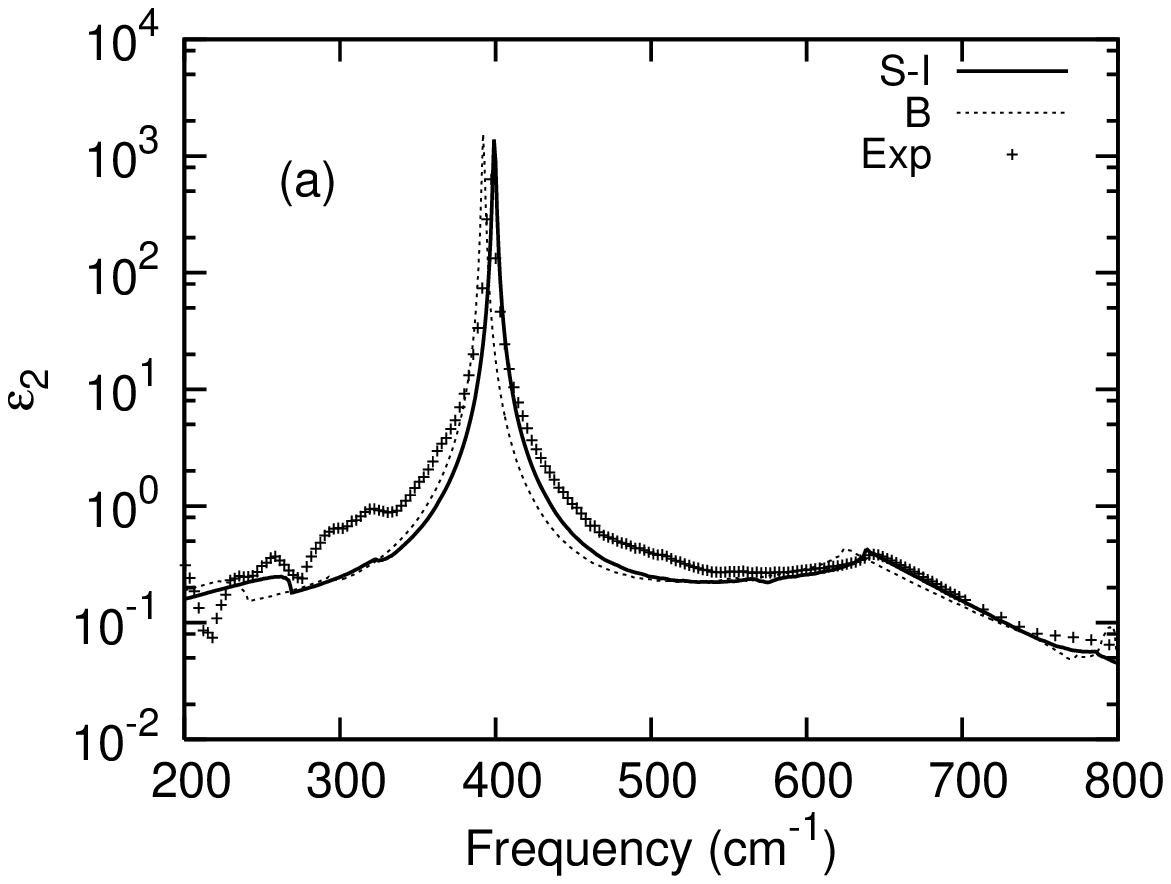}
\includegraphics[width=0.48\textwidth]{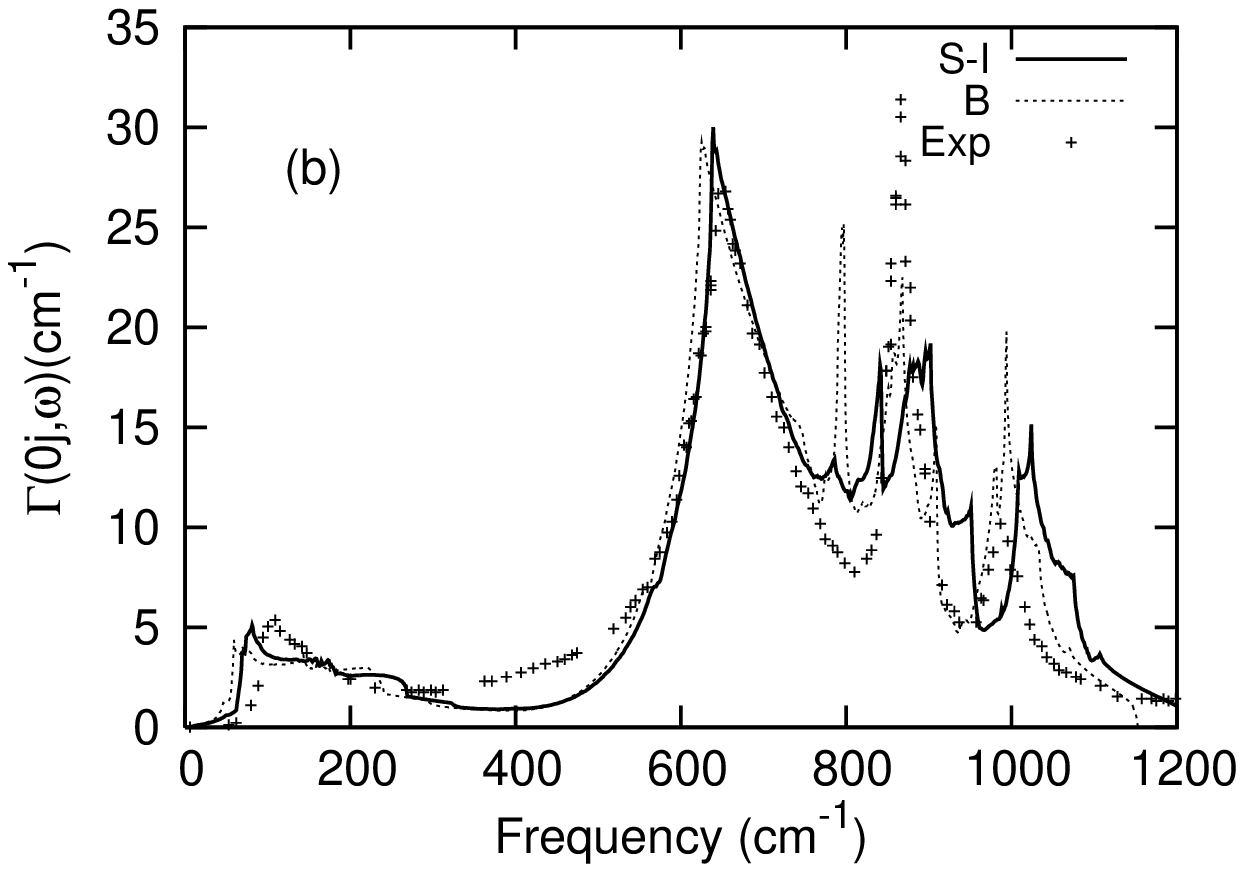}
\caption{Computed anharmonic properties compared with experimental data for
pure MgO. (a)~The imaginary part of dielectric function at 295~K; the experimental
data are the same as those in Fig.~\ref{fig:exp}(d). (b)~The imaginary part of
self energy at 295~K; the experimental data are digitized from
Ref.~\onlinecite{meneses}, which are fit to infared spectra based on a
semi-quantum dielectric function model.} \label{fig:anha}
\end{figure}

The anharmonic effects in pure MgO are shown in Fig.~\ref{fig:anha}. The
computational results and experimental values are quite close, especially near
640~cm$^{-1}$ which corresponds to TO+TA combination mode. It is not
surprising, since the procedure we followed was originally developed and worked
well for alkali-halide salts, which are similar to MgO. Below 800~cm$^{-1}$,
the rigid shell model S-I and breathing shell model B give almost identical
self-energies. The discrepancy in the high-frequency range indicates that the
dispersion relations from empirical models are less accurate for high-frequency
optical branches. The width of TO mode at the reststrahlen frequency
$\omega_{\rm TO}$ is less accurate, as $\Gamma_{\rm anharm}(0j,\omega_{\rm
TO})$ is intrinsically small and higher-order anharmonic effects become
important.\cite{bruce}
%
% Figure 3
%

\begin{figure}
\includegraphics[width=0.48\textwidth]{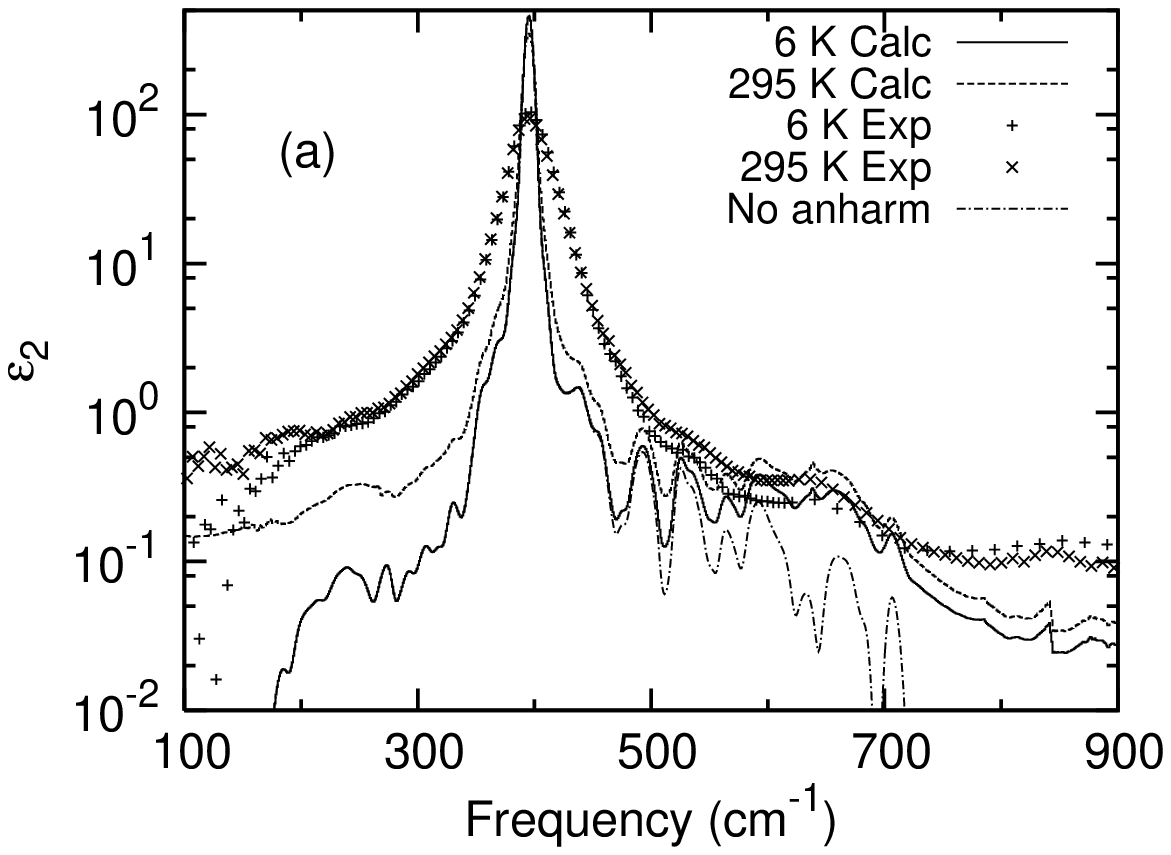}
\includegraphics[width=0.48\textwidth]{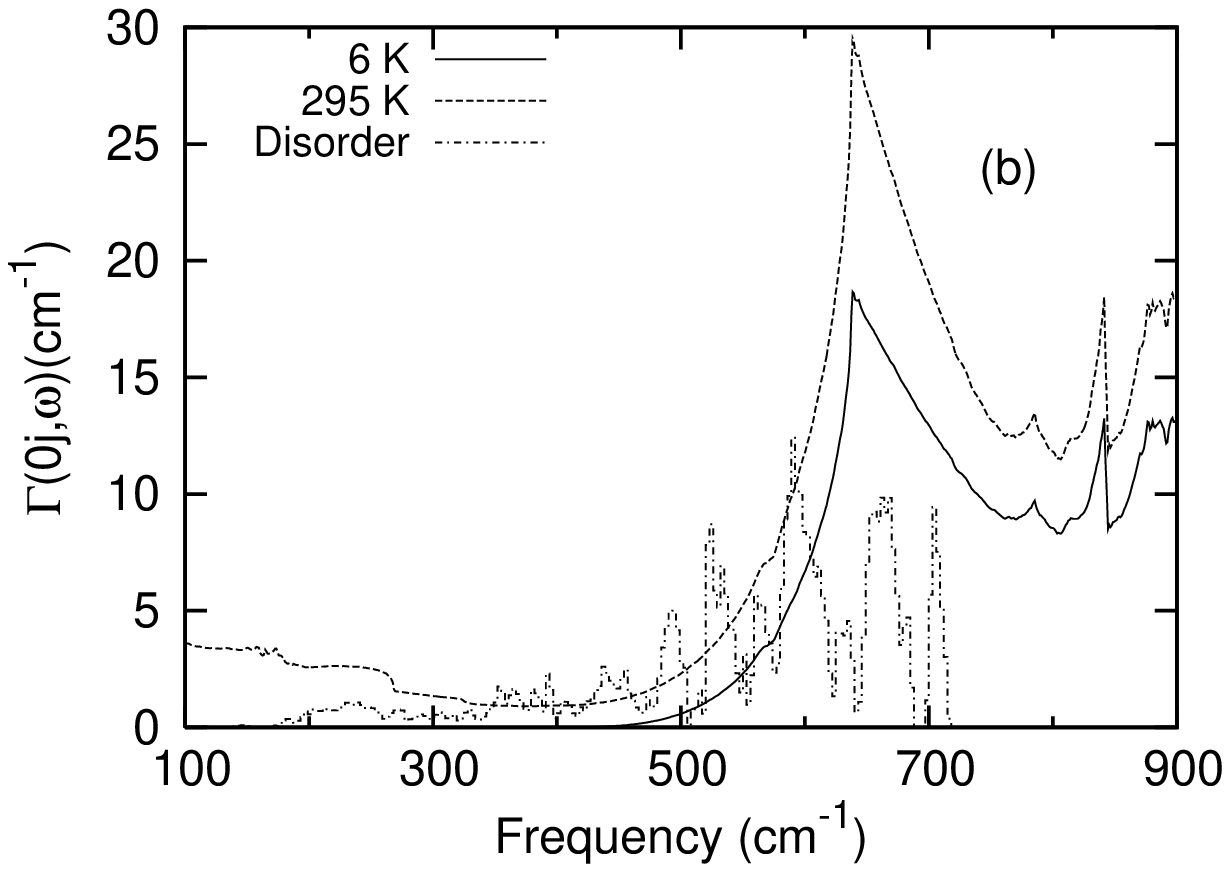}
\caption{The anharmonic and disordering scattering effects in Mg$_{1-x}$Fe$_x$O
for the 6\% Fe-doped sample. (a)~Imaginary part of the dielectric function. The
labels `6~K Calc' and `295~K Calc' denote the calculated curves, including both
disorder scattering and anharmonic interactions at the corresponding 
temperature. Experimental data are the same as those in Fig.~\ref{fig:exp}(e).
The label `No anharm' denotes the dielectric function calculated from disorder
scattering only.
(b)~Imaginary part of self energy. The labels `6~K' and `295~K' denote the
self-energies caused by anharmonic interaction at the corresponding temperature;
`disorder' denotes the self-energy due to disorder scattering, which is 
computed by histogram method where the bin size equals 1~cm$^{-1}$, then 
iteratively averaged 30 times. The total self-energies are the sum of these 
two pieces, and are used in calculating the `6~K Calc' and `295~K Calc' 
dielectric functions shown in (a).}
\label{fig:6percent}
\end{figure}

\begin{figure}
\includegraphics[width=0.48\textwidth]{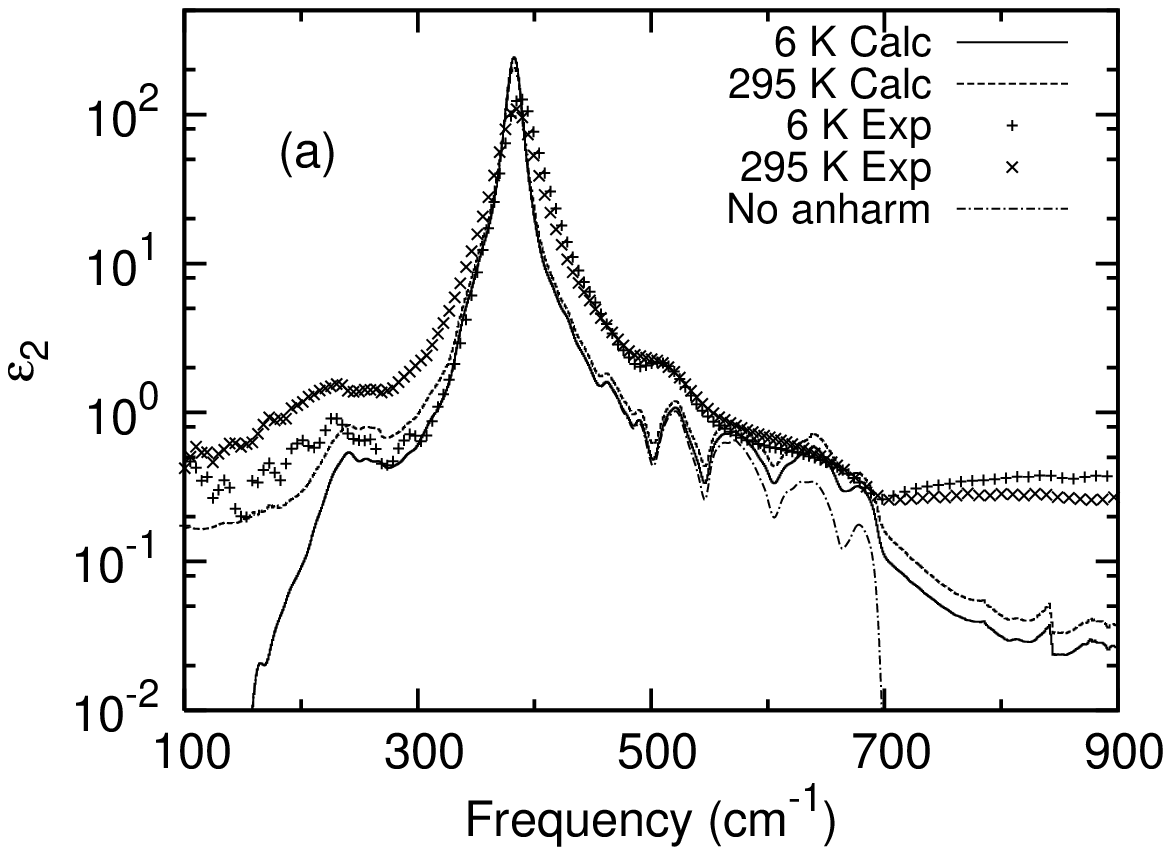}
\includegraphics[width=0.48\textwidth]{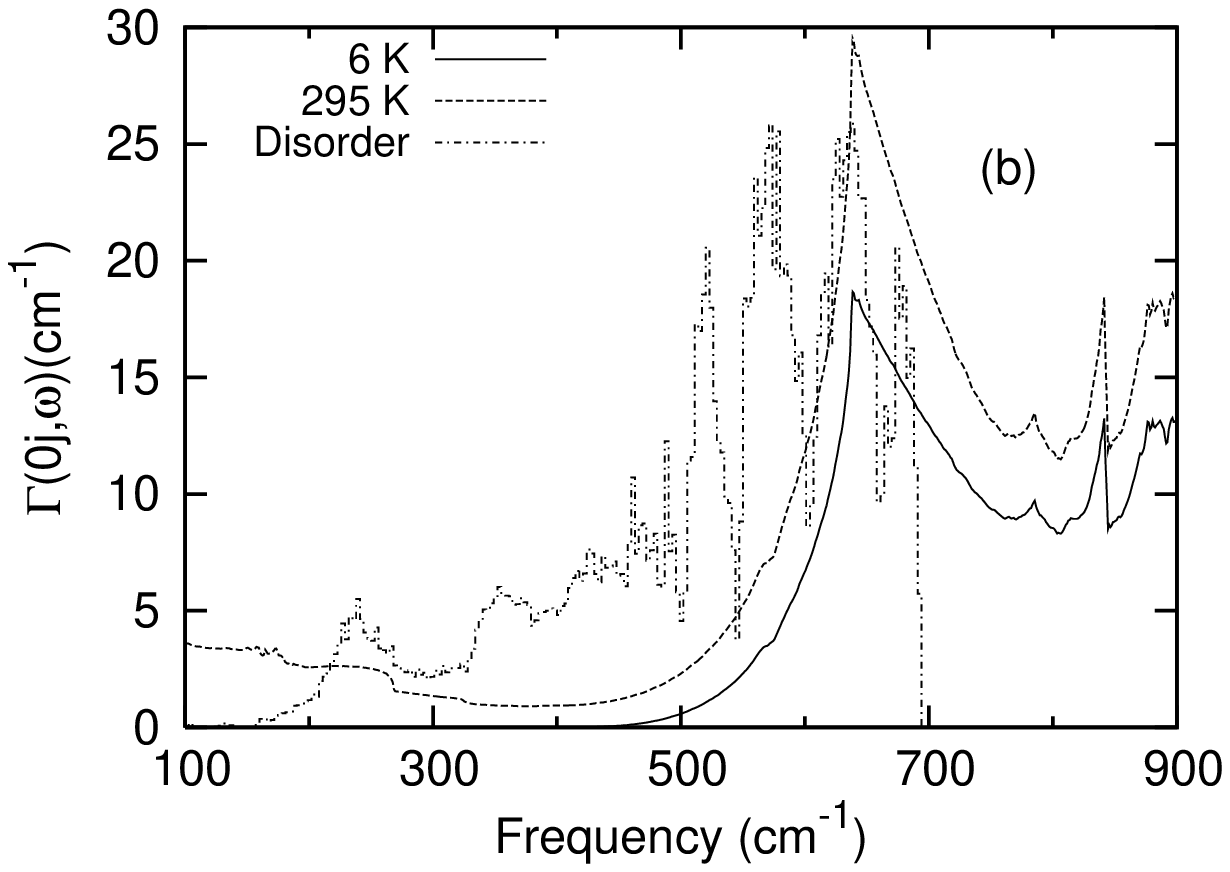}
\caption{The anharmonic and disordering scattering effects in Mg$_{1-x}$Fe$_x$O
for the 27\% Fe-doped sample. (a)~Imaginary part of the dielectric function;
(b)~Imaginary part of self energy. The computation procedure is the same as for
the 6\% Fe doping.}
\label{fig:27percent}
\end{figure}

Figures~\ref{fig:6percent} and \ref{fig:27percent} show how anharmonicity and
disorder scattering influence the dielectric function. For the 6\% sample it is
clear that the shoulder near 640~cm$^{-1}$ is caused by anharmonicity, while
the shoulder at about 520~cm$^{-1}$ is due to disorder scattering. Disorder
scattering becomes stronger for the 27\% sample and seems contributes to all
the shoulders. The shoulder caused by anharmonicity corresponds to a peak in
the two-phonon DOS, while shoulders caused by disorder scattering are related
to peaks in the one-phonon DOS.
%
% Figure 4
%

%
% Figure 5
%

\begin{figure}
\includegraphics[width=0.48\textwidth]{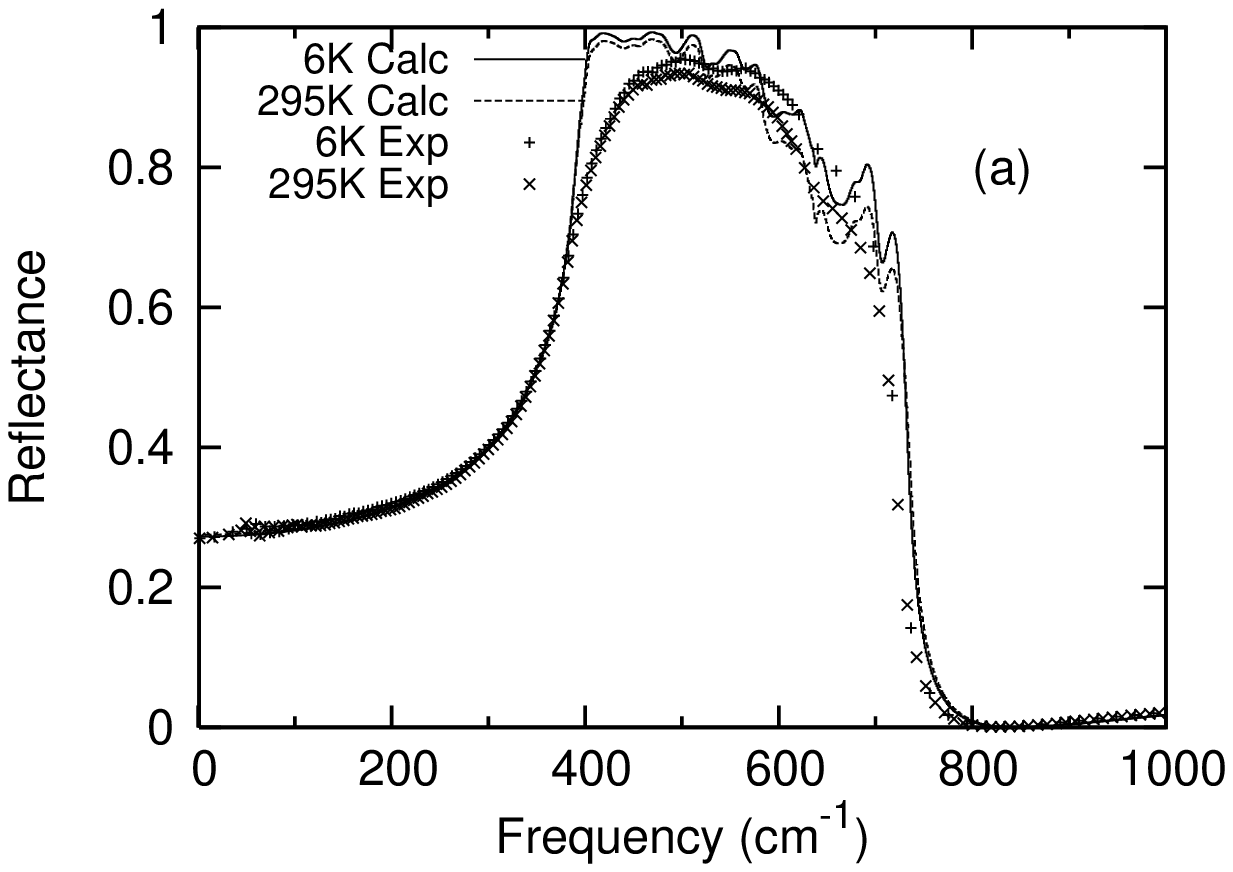}
\includegraphics[width=0.48\textwidth]{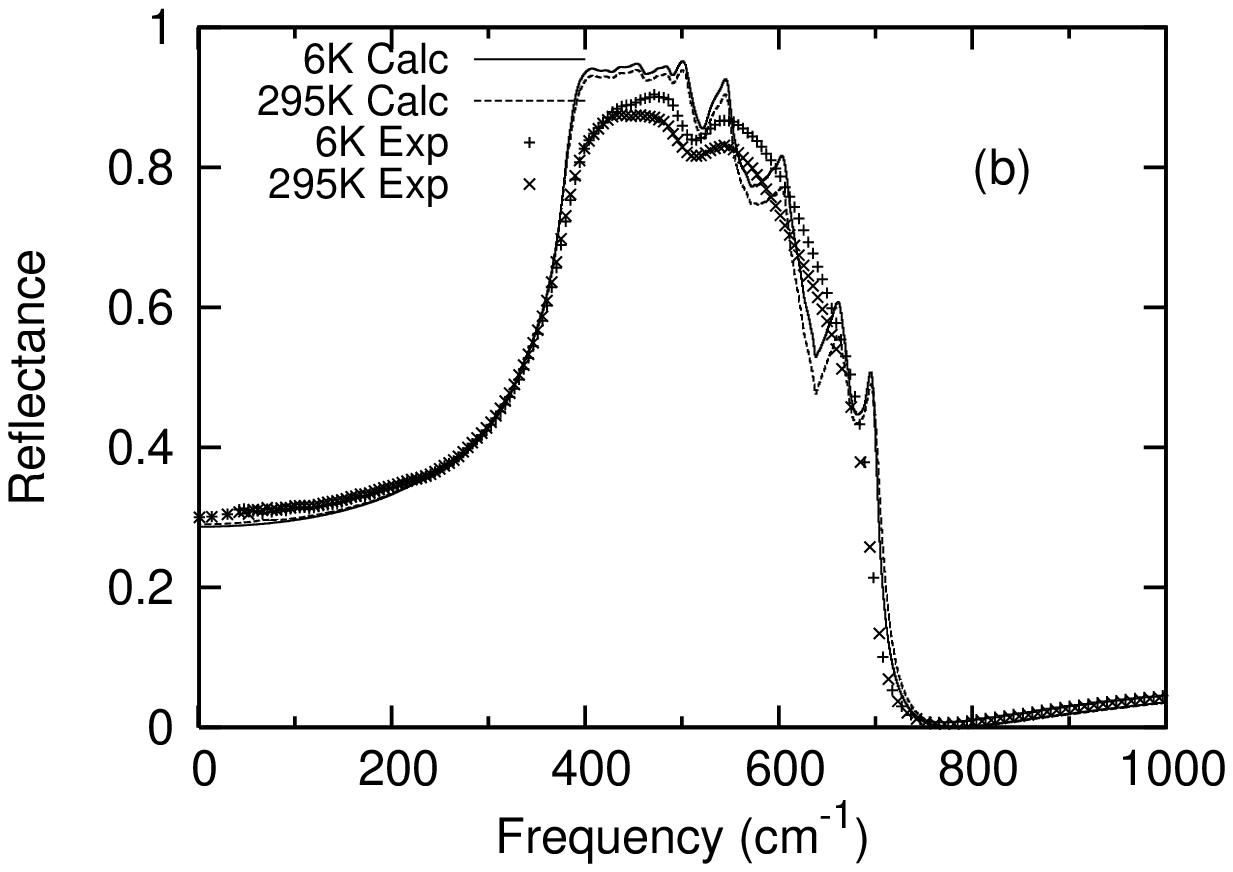}
\caption{The calculated infrared reflectance, compared with the experimental
data~(same as in Fig.~\ref{fig:exp}(b) and (c)) for Mg$_{1-x}$Fe$_x$O. 
(a)~6\% Fe doping; (b)~27\% Fe doping.}
\label{fig:reflectance}
\end{figure}

Figure~\ref{fig:reflectance} contains the reflectance computed from the
dielectric functions at 295 and 6~K shown in Figs.~\ref{fig:6percent} and
\ref{fig:27percent}. As in the case of pure MgO, the agreement between theory
and experiment is better in the region where the self-energy caused by
lowest-order pertubation is large. Near the reststrahlen frequency $\omega_{\rm
TO}$, the self-energy is smaller, and $R(\omega)$ is more sensitive to details.
Our model underestimates the broadening of the resonance, but correctly
identifies the sources of broadening.
%
% Figure 6
%

It is of interest to determine whether the disorder scattering is mainly due to
the differences in mass or in the inter-atomic potential. Thus we repeat the
above procedure with a model which only contains mass disorder, i.e. Fe is
treated as an isotope of Mg, its shell parameters are the same as Mg$^{2+}$ in 
S-I model. It turns out the most significant factor is $\epsilon_{\infty}$: For 
the isotope model~(S-I) $\epsilon_{\infty}$ is the same as pure MgO~(2.94), 
for S-II model $\epsilon_{\infty}$ increases to 3.05 for 6\%
Fe and 3.47 for 27\%, in reasonable agreement with the results shown in
Table~\ref{tab:exp}. Thus the LO frequency predicted from the isotope model is
larger than the experimental value. The differences in the inter-atomic
potentials change the relative strength of the self-energy, but in both cases
the self-energy spectra carry features of the one phonon DOS of pure MgO.

In addition to phonons, electronic transitions may also influence the infrared
dielectric properties of ferropericlase. Wong\cite{wong} measured the
far-infrared absorption spectra of iron-doped MgO. A line at 105~cm$^{-1}$ was
observed with a peak absorption coefficient of 1.5~cm$^{-1}$ and a width of
$\simeq 9$~cm$^{-1}$ at 20~K in a sample with 0.2\% Fe. This feature is
attributed to the transition $\Gamma_{\rm 5g}\rightarrow\Gamma_{\rm
3g},\Gamma_{\rm 4g}$ of MgO:~Fe$^{2+}$ at cubic sites. If we assume the
absorption coefficient is proportional to the impurity concentration, then we
can estimate the corresponding $\epsilon_2$ at 105~cm$^{-1}$ by
$\epsilon_2(\omega) = \frac{n\alpha(\omega)}{2\pi\omega}$, where $n$ is the
refractive index~(for pure MgO, $n \simeq 3.2$ at 105~cm$^{-1}$),
$\alpha(\omega)$ is the absorption coefficient at frequency $\omega$~(in units
of cm$^{-1}$). The value of $\epsilon_2$ is about 0.22 for 6\% Fe
concentration, 0.98 for 27\%. As the iron concentration increases, the
electronic transitions of Fe$^{2+}$ should show greater influence on the
far-infrared spectra of ferropericlase. In our measurement the spectra below
200~cm$^{-1}$ are complicated due to the presence of fringes, consequently we
can not confirm this tendency. Henning {\it et al.}\cite{henning} measured the
infrared reflectance of Fe$_{x}$Mg$_{1-x}$O for $x=1.0$, 0.9, 0.8, 0.7, 0.5 and
0.4 at room temperature. The $\epsilon_2$ curves reported in their paper do not show a monotonic rise in the far-infrared region as the iron concentration $x$
increases from 0.4 to 1.0, while they are all in the range of $6$-$10$ near
100~cm$^{-1}$. It is difficult to explain such large $\epsilon_2$ with lattice
vibrations alone, and the accuracy of these data has been
questioned.\cite{hofmeister03} Further experiments are needed to clarify this
issue.

%
% Conclusions
%
\section{Conclusions}
The infrared reflectance spectra of magnesium oxide and ferropericlase has
been measured at 295 and 6~K. It is found that $\epsilon_{\infty}$ increases as
Fe concentration increases, while the width of the TO modes remains the same in
the doped materials. We construct a theoretical model which includes both
disorder scattering and anharmonic phonon-phonon interactions. The model shows
fairly good agreement with the experiment in the regions where the lowest-order
perturbation is relatively large. Near the resonance, theory and experiment
both have smaller self-energies, which makes the reflectance quite sensitive
to the details. We do not know whether the disagreements with experiment in
the region are caused by neglect of higher order corrections, or by inaccuracy
of the underlying model. However, the model identifies the global features
reasonably well, and may provide a good basis for the study of phonon decay
needed for a theory of heat conductivity.

%
% Acknowledgements

\acknowledgments
This work was supported by the Office of Science, U.S. Department of Energy,
under Contract No.~DE-AC02-98CH10886. SDJ is supported by NSF EAR-0721449.

\bibliography{citation}
\end{document}